\documentclass[preprint,prb,showpacs,preprintnumbers,amsmath,amssymb]{revtex4}

\usepackage{graphicx}

\begin{document}

\title{Mass renormalization in the band width-controlled
Mott-Hubbard systems SrVO$_3$ and CaVO$_3$ studied by
angle-resolved photoemission spectroscopy}

\author{T. Yoshida$^1$,
 M. Hashimoto$^1$, T. Takizawa$^1$,
A. Fujimori$^1$, M. Kubota$^2$, K. Ono$^2$ and H. Eisaki$^3$}
\affiliation{$^1$Department of Physics, University of Tokyo,
Bunkyo-ku, Tokyo 113-0033, Japan} \affiliation{$^2$KEK, Photon
Factory, Tsukuba, Ibaraki 305-0801,
Japan}\affiliation{$^3$National Institute of Advanced Industrial
Science and Technology, Tsukuba 305-8568, Japan}
\date{\today}

\begin{abstract}
Ca$_{1-x}$Sr$_x$VO$_3$ is a Mott-Hubbard-type correlated electron
system whose bandwidth can be varied by the V-O-V bond angle, but
the actual effect of bandwidth control on the electronic structure
has been controversial in previous photoemission experiments. In
this work, band dispersions and Fermi surfaces of SrVO$_3$ and
CaVO$_3$ are studied by angle-resolved photoemission spectroscopy.
Near the Fermi level ($E_F$), three bands forming cylindrical
Fermi surfaces derived from the three V 3$d$ $t_{2g}$ orbitals
have been observed. The observed band widths for both compounds
are almost half of those predicted by local-density-approximation
band-structure calculation, confirming mass renormalization caused
by electron correlation. It has been clearly demonstrated that the
width of the $d$ band in CaVO$_3$ is narrower than that in
SrVO$_3$, qualitatively consistent with the result of
band-structure calculation. Roles of the orthorhombic lattice
distortion and electron correlation in the observed band narrowing
are discussed.
\end{abstract}

\maketitle
The complex nature of correlated electrons in transition-metal
oxides (TMO's) causes various interesting phenomena including
high-$T_c$ superconductivity and colossal magnetoresistance
\cite{RMPfujimori}. Light transition-metal oxides such as
perovskite-type Ti and V oxides are ideal systems to study the
fundamental physics of electron correlation  because they are
prototypical Mott-Hubbard-type systems, in which the O 2$p$ band
is located below the transition-metal 3$d$ band, in the framework
of the Zaanen-Sawatzky-Allen classification scheme.\cite{ZSA}
Those systems have a relatively small number of electrons in the
degenerate $t_{2g}$ bands, and their electronic properties have
been modeled by the Hubbard model without explicit consideration
of the oxygen $p$ orbitals. In the Mott-Hubbard regime, the MIT
occurs when the ratio of the on-site Coulomb repulsion ($U$) to
the one-electron band width ($W$) exceeds a critical value
$(U/W)_c$. In those light TMO's, the electronic properties are
those of normal Fermi liquids on the metallic side, while they are
Mott insulators (with orbital ordering) on the insulating side of
the MIT. Electron correlation effects and metal-insulator
transitions (MITs) in light TMO's have been described based on
theoretical predictions of dynamical mean-field theory (DMFT) for
the Hubbard model \cite{Kotliar}.

 To address the nature of electron correlation in light TMO's experimentally,
photoemission spectroscopy has provided rich information. In the
photoemission spectra of the perovskite-type Ti and V
oxides,\cite{inouePES, Morikawa, Fujimori, YoshidaLSTO} the
coherent part around the Fermi level ($E_F$) corresponding to
band-like electronic excitations and incoherent part 1-2 eV away
from $E_F$ corresponding to atomic-like excitations or the remnant
of the lower Hubbard band (LHB) have been observed. DMFT has
predicted that the effective mass of conduction electron is
enhanced concomitant with decreasing spectral weight in the
coherent part \cite{Kotliar} and that spectral weight transfer
occurs from the coherent part to incoherent part as the system
approaches an MIT from the metallic side. For example, in the
filling-control Mott-Hubbard system La$_{1-x}$Sr$_x$TiO$_3$
(LSTO), a critical mass enhancement occurs toward the MIT
according to the electronic specific heats $\gamma$ and the
magnetic susceptibility $\chi$.\cite{Kumagai} The doping
dependence of photoemission spectral weight and of the bandwidth
of the coherent part indeed reflects the behaviors of $\gamma$ and
$\chi$ as predicted by DMFT.\cite{YoshidaLSTO} On the other hand,
the effective mass enhancement and spectral weight transfer in the
bandwidth-control Mott-Hubbard system Ca$_{1-x}$Sr$_x$VO$_3$
(CSVO) have been controversial. In CSVO, the V-O-V angle varies
with $x$ and the band width decreases with decreasing $x$. Early
photoemission results have shown that, with decreasing $x$, i.e.,
with decreasing bandwidth, spectral weight is transferred from the
coherent part to the incoherent part \cite{inouePES} in a dramatic
way compared to the moderate enhancement of
$\gamma$.\cite{inouePRB} In contrast, according to a
bulk-sensitive photoemission study using soft x-rays, neither
appreciable spectral weight transfer nor appreciable band
narrowing has been observed for SrVO$_3$ and
CaVO$_3$.\cite{Sekiyama} On the other hand, another bulk-sensitive
 photoemission study using a laser has revealed the suppression of spectral
weight near $E_F$ in going from SrVO$_3$ to CaVO$_3$.\cite{Eguchi}
Thus, the difference in the electronic structure between SrVO$_3$
and CaVO$_3$ still remains unclear. To reconcile the discrepancies
between the different experiments caused by overlapping surface
and bulk signals, angle-resolved photoemission spectroscopy
(ARPES) turned out to be a powerful method, as demonstrated by the
successful ARPES observation of band dispersions and Fermi
surfaces in SrVO$_3$.\cite{YoshidaSVO} In that study, mass
renormalization which is consistent with $\gamma$ and reflects the
bulk electronic properties could be identified. The result has
been confirmed by a subsequent ARPES study using epitaxially grown
thin films of SrVO$_3$ with high surface quality.\cite{Takizawa}
However, how the electronic structure changes with bandwidth
control in CSVO has not been studied using ARPES. In order to
address this issue, in the present work, we have performed an
ARPES study of SrVO$_3$ and CaVO$_3$. In going from SrVO$_3$ to
CaVO$_3$, one would expect that the observed band dispersions
reflect band narrowing effects both due to orthorhombic lattice
distortion and electron correlation. In this work, we have clearly
observed the narrower band width in CaVO$_3$ than that in
SrVO$_3$, which is quantitatively consistent with the specific
heat coefficient $\gamma$. By comparing the ARPES results with the
results of band-structure calculations \cite{PavariniNJP,
Nekrasov}, the effects of the orthorhombic lattice distortion and
electron correlation on the observed mass renormalization are
discussed.

ARPES measurements were performed at beamline 28A of Photon
Factory with a Scienta SES-2002 electron analyzer. The typical
energy and angular resolutions were about 30 meV and 0.3 degree,
respectively. Single crystals of SrVO$_3$ and CaVO$_3$ were grown
by the travelling-solvent floating zone method. Samples were first
aligned \textit{ex situ} using Laue diffraction, cleaved
\textit{in situ} along the cubic (100) surface at 20 K and
measured at the same temperature in a pressure better than $1
\times 10^{-10}$ Torr. We used circularly polarized photons with
energies from $h\nu$=47 to 100 eV. In this paper, the electron
momentum is expressed in units of $\pi/a$, where $a = 3.84$\AA
\space (3.76\AA) for SrVO$_3$ (CaVO$_3$) is the cubic lattice
constant corresponding to the V-V distance. $k_x$ and $k_y$ are
the momenta parallel to the cleavage plane and $k_z$ is the
momentum perpendicular to the plane.

\begin{figure}
\includegraphics[width=16cm]{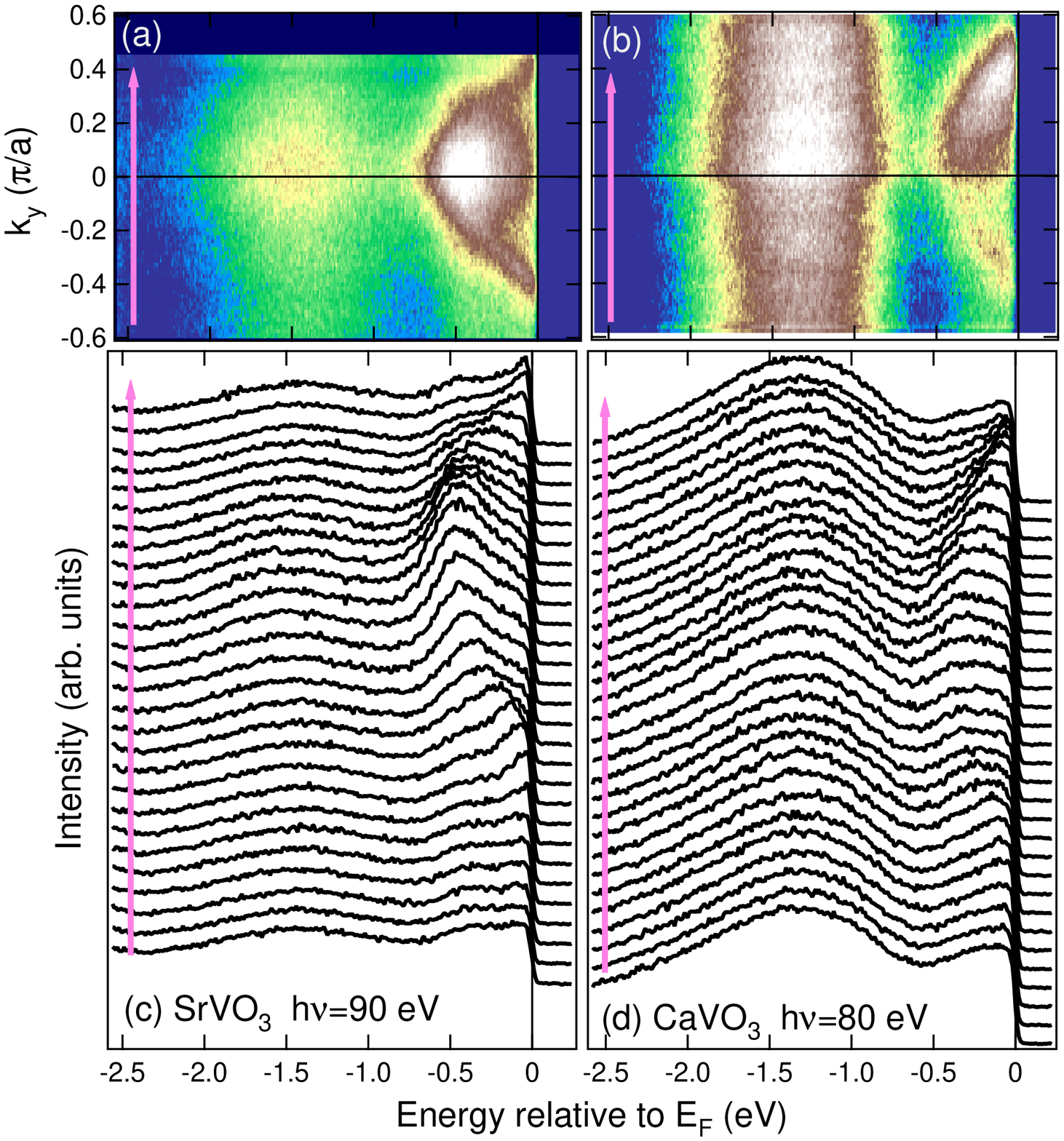}
\caption{\label{incoherent}(Color online) ARPES spectra of
SrVO$_3$ and CaVO$_3$ along the $k_y$ direction with $k_x$=0 in
the first Brillouin zone. (a)(b) Intensity plot of SrVO$_3$ and
CaVO$_3$ in the $E$-$k_y$ plane. (c)(d) Corresponding EDC's. The
dispersive feature within $\sim$0.7 eV of $E_F$ is the coherent
part, while the broad feature centered around 1.3-1.5 eV below
$E_F$ is the incoherent part.}
\end{figure}

First, we show energy distribution curves (EDCs) of SrVO$_3$ and
CaVO$_3$ for $k_x$=0 with various $k_y$'s in Fig.
\ref{incoherent}. The coherent part within $\sim$ 0.7 eV of the
Fermi level ($E_F$) shows a clear dispersive feature corresponding
to the calculated band structure. The incoherent part, which
reflects electron correlation, centered at $\sim$ -1.5 eV and
$\sim$ -1.3 eV for SrVO$_3$ and CaVO$_3$, respectively, shows
appreciable momentum dependent intensities as seen in Figs.
\ref{incoherent}(a) and \ref{incoherent}(b). The intensity is
stronger within the Fermi surface ($k_y<k_F$), consistent with the
previous ARPES study on thin films and the results of DMFT
calculation.\cite{Takizawa}

\begin{figure}
\includegraphics[width=16cm]{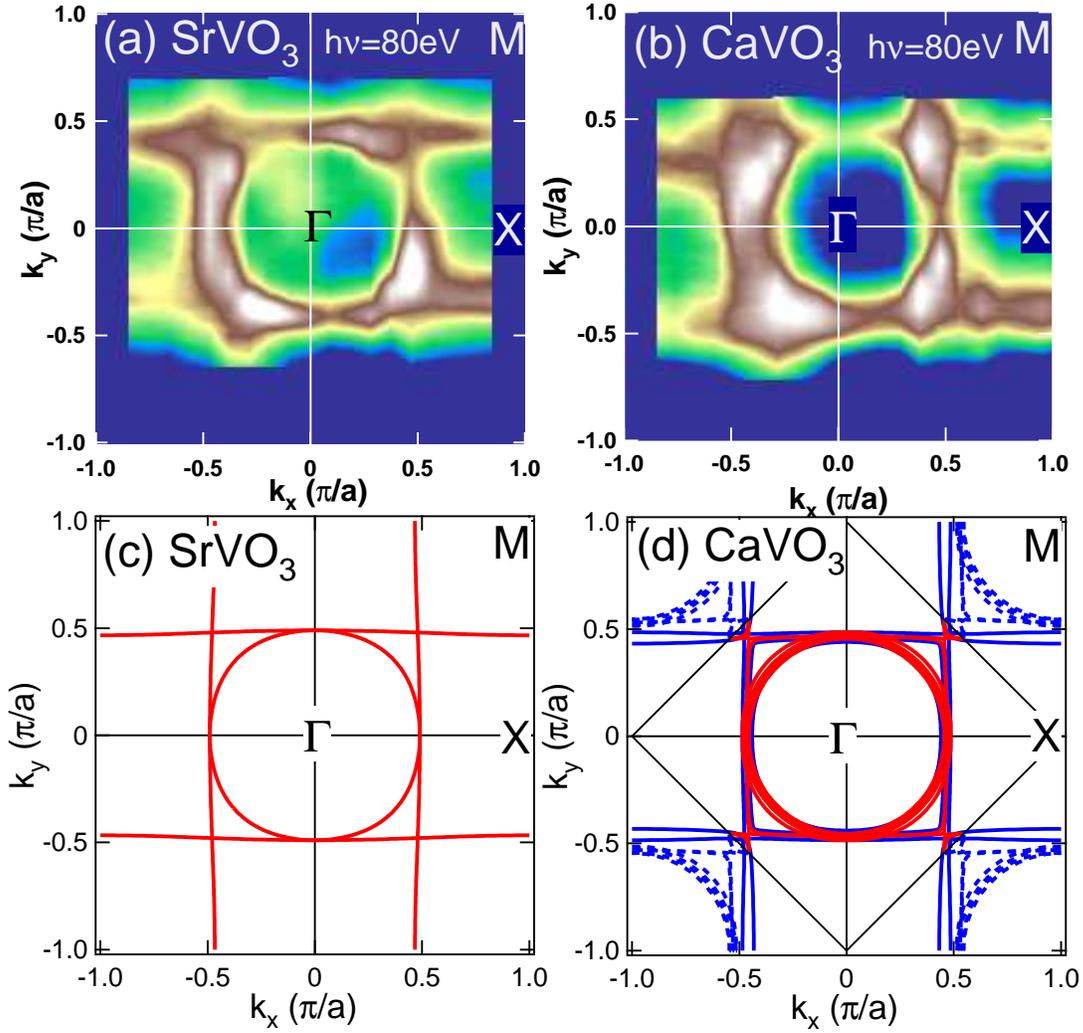}
\caption{\label{nkMapping}(Color online) Spectral weight mapping
at $E_F$ for SrVO$_3$ (a) and CaVO$_3$ (b) using $h\nu$=80 eV.
Spectral weight has been integrated within 20 meV of $E_F$. Note
that the mapping is a projection on the $k_x$-$k_y$ plane. (c)(d)
Fermi surface cross-sections by the $k_z$=0 plane (red curves)
predicted by band-structure calculation \cite{PavariniNJP}. For
CaVO$_3$, Fermi surface cross-sections by the $k_y$=0 and $k_x$=0
planes and folded Fermi surfaces due to the orthorhombic
distortion are indicated by overlapping blue solid and dashed
curves, respectively.}
\end{figure}

Spectral weight at $E_F$ in the first Brillouin zone (BZ) is
mapped in $k_x$-$k_y$ space for SrVO$_3$ and CaVO$_3$ in Figs.
\ref{nkMapping} (a) and (b), respectively, for photon energy
$h\nu$=80 eV. Fermi surfaces for SrVO$_3$ and CaVO$_3$ predicted
by band-structure calculation \cite{PavariniNJP} are shown in
Figs. \ref{nkMapping} (c) and (d), respectively. For SrVO$_3$, the
overall feature of the mapped Fermi surface is in good agreement
with the band-structure calculation. The observed spectral weight
distribution indicates the cylindrical Fermi surfaces consisting
of the three $t_{2g}$ orbitals, the $d_{xy}$, $d_{yz}$ and
$d_{zx}$ orbitals, of vanadium. Particularly, we have observed a
cross-section or a projection of the cylindrical Fermi surface
derived from the $d_{yz}$ orbital and that from the $d_{zx}$
orbital. These Fermi surfaces are extended along the $k_x$ and
$k_y$ directions and were not clearly observed in the previous
study.\cite{YoshidaSVO} The result of the spectral weight mapping
for CaVO$_3$ is similar to that of SrVO$_3$, but the momentum
distribution is generally broader than SrVO$_3$ because the
crystal structure of CaVO$_3$ is orthorhombic and band folding
occurs due to the quadrupling of the unit cell as shown in
Fig.\ref{nkMapping} (d). Since the cleaved surface may contain
$ab$, $bc$, and $ac$ planes, the observed ARPES spectra would be a
superposition of the dispersions from the $ab$, $bc$, and $ac$
surfaces. Such folded bands of different orientations are not
resolved in the present spectra, and would give rise to the broad
spectral weight distribution compared to that in SrVO$_3$.

\begin{figure}
\includegraphics[width=16cm]{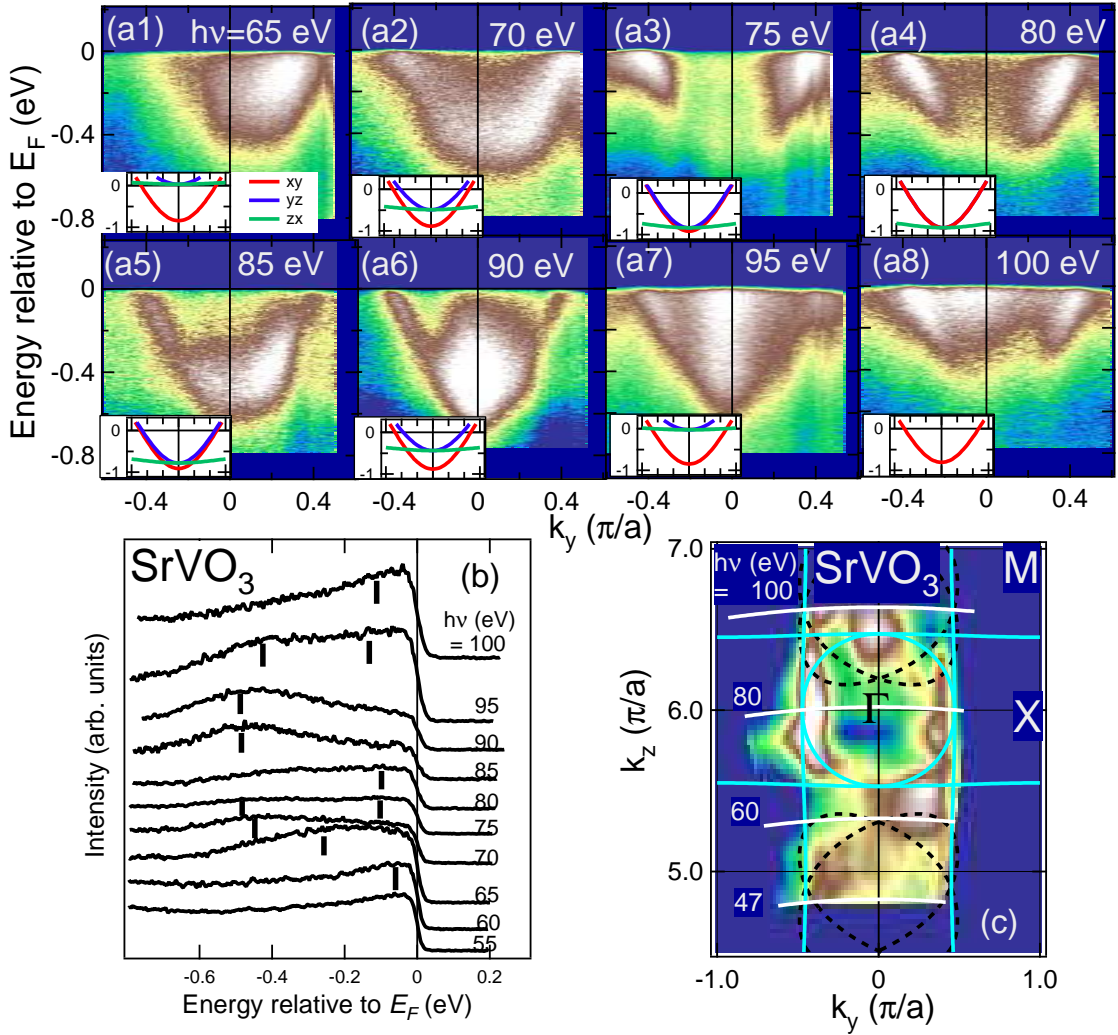}
\caption{\label{hnSVO}(Color online) ARPES spectra of SrVO$_3$ in
the $E$-$k_y$ plane for various photon energies corresponding to
various $k_z$'s. (a1)-(a8): Photon energy dependence of the ARPES
spectra near the normal emission. Insets are band dispersions for
corresponding cuts predicted by band-structure calculation.
\cite{PavariniNJP} (b) Normal emission spectra. (c) Intensity at
$E_F$ mapped in the $k_y$-$k_z$ plane. Fermi surfaces predicted by
band-structure calculation \cite{PavariniNJP} are shown by blue
curves. Black dotted curves indicate the $d_{yz}$ Fermi surfaces
from secondary cones. }
\end{figure}

\begin{figure}
\includegraphics[width=16cm]{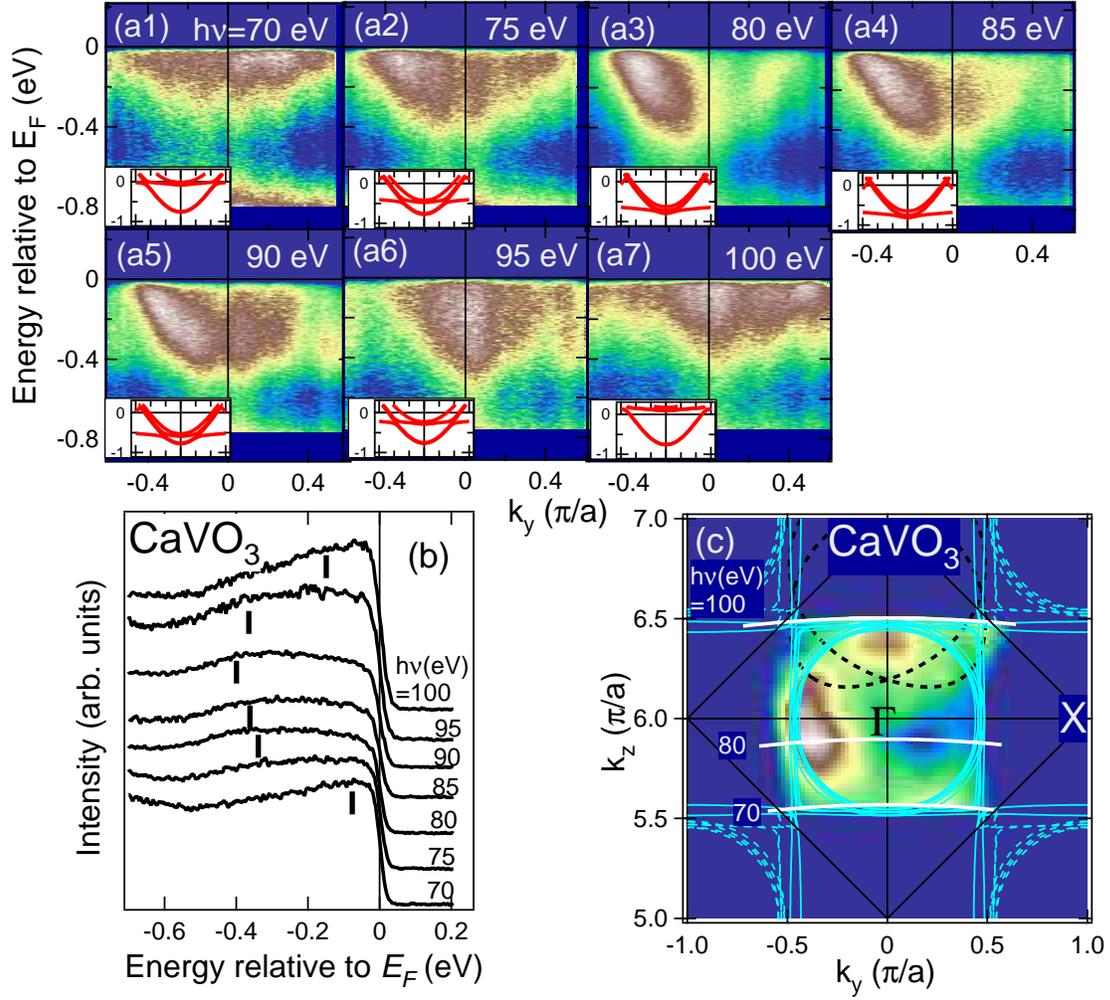}
\caption{\label{hnCVO}(Color online) ARPES spectra of CaVO$_3$ for
various photon energies corresponding to various $k_z$'s.
(a1)-(a7): Photon energy dependence of the ARPES spectra around
the normal emission. Insets are band dispersions for corresponding
cuts predicted by band-structure calculation. \cite{PavariniNJP}
(b) Normal emission spectra. (c) Intensity at $E_F$ mapped in
$k_y$-$k_z$ space. Fermi surfaces predicted by band-structure
calculation \cite{PavariniNJP} are shown by blue curves. Black
dotted lines indicate $d_{yz}$ Fermi surfaces from secondary
cones.}
\end{figure}

Figures \ref{hnSVO}(a) and \ref{hnCVO}(a) show the photon energy
dependence of the ARPES spectra of SrVO$_3$ and CaVO$_3$,
respectively, for $k_x$=0 cuts along the $k_y$ direction. The
asymmetry of the ARPES intensity with respect to $k_y$=0 is due to
matrix-element effects of the circular polarized light. The
observed parabolic band is the $d_{xy}$ band, which has nearly
two-dimensional electronic structure in the $k_x$-$k_y$ plane
parallel to the sample surface. Spectral features of CaVO$_3$ are
broader than those of SrVO$_3$, which may be due to the
orthorhombic crystal distortion of CaVO$_3$. For SrVO$_3$, we have
observed enhanced spectral weight at -0.4 eV for $h\nu\sim$ 90 eV
and this enhanced part moves toward the Fermi level with
increasing photon energy. This enhancement is due to the overlap
of the bottom $d_{yz, zx}$ band because the matrix element of the
$d_{xy}$ orbital is suppressed around $k_x$=$k_y$=0. Therefore,
the shift of the intensity with photon energy indicates the energy
dispersion of the $d_{yz, zx}$ bands along the $k_z$ direction.
This behavior is also predicted by the band calculation
\cite{PavariniNJP} as shown in the insets. In the case of
CaVO$_3$, the intensity of the $d_{yz, zx}$ bands are not so
strong as in SrVO$_3$. Nevertheless, the flat dispersions near
$E_F$ corresponding to the $d_{yz, zx}$ bands are observed around
$h\nu$= 70 eV and 100 eV, indicating the Fermi surface crossing of
these bands along the $k_z$ direction. Figures \ref{hnSVO} (b) and
\ref{hnCVO} (b) are normal emission spectra for various photon
energies. We have also mapped ARPES intensities at $E_F$ in the
$k_y$-$k_z$ plane in order to reveal the cross-section of the
$d_{yz}$ FS, as shown in Figs. \ref{hnSVO} (c) and \ref{hnCVO}
(c). Here, the $k_z$ values have been obtained from $h\nu$'s
assuming the inner potential of $V_0$=17 eV. These mapping
patterns for both samples qualitatively agree with the Fermi
surfaces obtained from the band-structure calculation as
demonstrated in Figs. \ref{hnSVO} (c) and \ref{hnCVO}(c). Note
that, in the previous ARPES study, signals from the $d_{yz, zx}$
bands were unclear, probably due to surface
effects,\cite{YoshidaSVO} while we have clearly observed these
bands in the present results. We should also remark that there are
structures with weak intensity which are not predicted by the band
calculation as shown in Fig. \ref{hnSVO}(c). This may represent
photoelectrons from secondary cones.\cite{Mahan} Assuming that the
emitted electrons receive the in-plane reciprocal vector
$\mathbf{G}$=($\pm2\pi$,$\pm2\pi$), we have simulated the $d_{yz}$
Fermi surface from secondary cones and explained the observed
intensity distribution, as indicated by dashed curves in Figs.
\ref{hnSVO}(c) and \ref{hnCVO}(c).

\begin{figure}
\includegraphics[width=16cm]{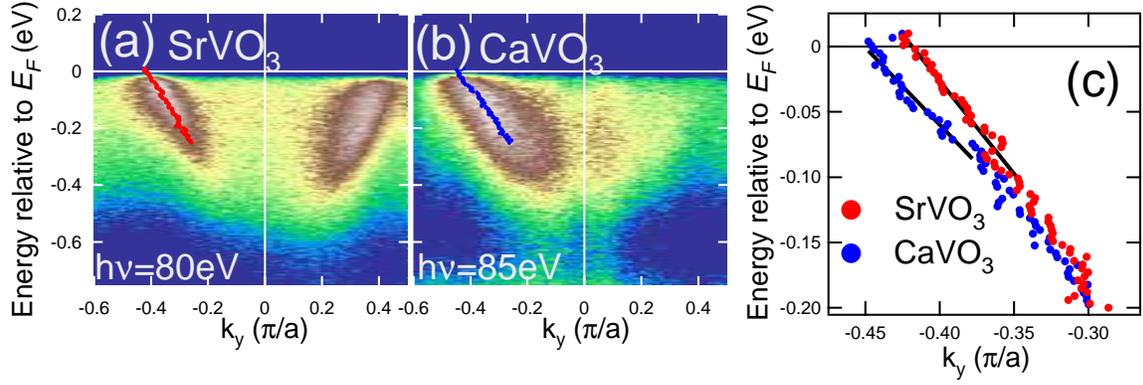}
\caption{\label{EDC_Ek}(Color online) Comparison of band
dispersions in SrVO$_3$ and CaVO$_3$. ARPES spectra of SrVO$_3$
(a) and CaVO$_3$ (b) along the $\Gamma$-X lines. The observed band
is the $d_{xy}$ band. (c) Quasi-particle dispersions determined by
the MDC peak positions.}
\end{figure}

In order to investigate electron correlation effects for both
samples, we have determined the QP dispersion near $\Gamma$ point
($h\nu\sim$ 80 eV for SrVO$_3$ and $h\nu\sim$ 85 eV for CaVO$_3$).
Figures \ref{EDC_Ek}(a) and \ref{EDC_Ek}(b) show the same ARPES
spectra as Fig. 3 (a4) and Fig. 4 (a4), respectively, with QP
dispersions determined by the peak positions of momentum
distribution curves (MDCs). The QP dispersions for both compounds
are compared in Fig. \ref{EDC_Ek}(c). From the slope of the
dispersions within $\sim$ 0.1 eV from the $E_F$, we have deduced
the Fermi velocity $v_F$ $\sim$ 1.7 and $\sim$ 1.4 eV\AA\space for
SrVO$_3$ and CaVO$_3$, respectively. This corresponds to the
observation that the binding energy of the bottom of the $d_{yz}$
band for CaVO$_3$ is $\sim$ 0.4 eV [Fig. \ref{hnCVO}(b)], while it
is $\sim$ 0.5 eV for SrVO$_3$ [Fig. \ref{hnSVO}(b)]. Assuming the
two-dimensional cylindrical Fermi surfaces for the three $t_{2g}$
orbitals, the electronic specific heat coefficient $\gamma$ has
been estimated from the observed $v_F$ to be $\gamma \sim$ 7.5 and
$\sim$ 9.0 mJ mol$^{-1}$ K$^{-2}$ for SrVO$_3$ and CaVO$_3$,
respectively, which are close to the experimental values of 8.18
(SrVO$_3$) and 9.25 (CaVO$_3$) mJ
mol$^{-1}$K$^{-2}$.\cite{inouePRB} Therefore, the observed mass
renormalization in the QP dispersions are quantitatively
consistent with the electronic specific heat coefficient $\gamma$
within the Fermi liquid picture. From these results, it is
experimentally confirmed that CaVO$_3$ has a narrower QP bandwidth
than SrVO$_3$, while this has been unclear in the previous
photoemission studies \cite{inouePES, Sekiyama, Eguchi}. Now, let
us discuss possible mechanisms of the band narrowing in these
materials. The observed band widths for both SrVO$_3$ to CaVO$_3$
are nearly half of those predicted by the band-structure
calculation. This can be attributed to electron correlation as
LDA+DMFT calculations explain the mass renormalization by a factor
of $\sim$2 if a moderate Coulomb interaction $U$ is
assumed.\cite{Pavarini}

 In going from SrVO$_3$ to CaVO$_3$, the additional band narrowing
has been observed as described above. The observed Fermi velocity
indicates an increase in the effective mass $m^*$ up to 18 \% .
This band narrowing may be interpreted in terms of either the
orthorhombic lattice distortion or electron correlation or both of
them. In the simple tight-binding description of the band
structure of the perovskite-type oxides, the effective hopping
parameter between neighboring $d$ orbitals is proportional to
$\cos^2\theta$, where $\theta$ is the V-O-V bond angle. In
SrVO$_3$ and CaVO$_3$, $\cos^2\theta$ is $\sim$ 1 ($\theta\sim$
180 $^\circ$) and $\cos^2\theta\sim$ 0.88 ($\theta\sim$ 160
$^\circ$), respectively, yielding a band narrowing by $\sim$12\%.
The LDA calculation by Pavarini \textit{et al}\cite{PavariniNJP}
has predicted the narrowing by $\sim 16\%$. However, another LDA
calculation by Nekrasov \textit{et al}\cite{Nekrasov} indicates
band narrowing only by $\sim$4\%, and this small narrowing is
explained by the increasing hopping parameter between
nearest-neighbor $d$ orbitals in CaVO$_3$. If we employ the LDA
calculation by Nekrasov \textit{et al.}\cite{Nekrasov}, i.e.
assuming that the LDA band mass $m_b$ increases by $\sim$4\%, the
present results indicate that $m^*/m_b$ increases by $\sim$10\% in
going from SrVO$_3$ to CaVO$_3$, consistent with the scenario that
$U/W$ increases and hence enhances $m^*/m_b$ through electron
correlation in going from SrVO$_3$ to CaVO$_3$. If we employ the
LDA calculation by Pavarini \textit{et al}, on the other hand, the
present results indicate that the electron mass enhancement
factors $m^*/m_b$ for both compounds are nearly the same between
SrVO$_3$ and CaVO$_3$ and therefore that there are no appreciable
difference in electron correlation strengths in both compounds. If
this interpretation is correct, $U/W$ should not increase
appreciably in going from SrVO$_3$ to CaVO$_3$. One possible
scenario for the nearly constant $U/W$ is that the on-site Coulomb
interaction $U$ is reduced by hybridization between the V 3$d$
orbitals and the $d$ orbitals of the A-site cation in going from
SrVO$_3$ to CaVO$_3$. If the orthorhombic distortion in CaVO$_3$
enhances the hybridization between these orbitals, the net density
of 3$d$ electrons at the V site may decrease, resulting in the
reduction of $U$. In fact, the binding energy of the incoherent
peak, which is approximately $\sim U/2$ according to the DMFT
results, decreases from $\sim$1.5 eV in SrVO$_3$ to $\sim$1.3 eV
in CaVO$_3$ as shown in Fig. \ref{incoherent}.

In conclusion, we have studied the energy dispersions and the
Fermi surfaces of the three-dimensional Mott-Hubbard systems
SrVO$_3$ and CaVO$_3$ by ARPES. The observed band widths for both
samples are almost half of those predicted by the band-structure
calculation, consistent with the DMFT calculation \cite{Pavarini}
where the mass renormalization is caused by electron correlation.
We have confirmed that the width of the V 3$d$ band indeed
decreases by $\sim 20\%$ in going from SrVO$_3$ to CaVO$_3$. The
observed mass renormalization in the $d$ band near $E_F$ can
explain the moderate mass enhancement in $\gamma$ from SrVO$_3$ to
CaVO$_3$. This band narrowing can be explained by the orthorhombic
distortion and possibly by additional increase of correlation
strength caused by the increase in $U/W$.

We are grateful to M. Rozenberg for enlightening discussions and
N. Kamakura for technical support. This work was supported by a
Grant-in-Aid for Scientific Research (No. 19204037) from the Japan
Society for the Promotion of Science (JSPS). This work was done
under the approval of the Photon Factory Program Advisory
Committee (Proposal No. 2006S2-001).

\bibliography{CVO}

\end{document}